\documentclass[openacc]{rsproca_new}

\usepackage{graphicx}
\usepackage{epstopdf}
\usepackage{epsfig}
\usepackage{amssymb,amsmath,stmaryrd,tabularx}
\usepackage[utf8]{inputenc}
\usepackage[center]{subfigure}
\usepackage{amssymb}
\usepackage{cleveref}
\usepackage[dvipsnames]{xcolor}
\usepackage[normalem]{ulem}
\usepackage{bm}
\usepackage{appendix}
\usepackage{amsmath}
\usepackage{empheq}
\usepackage{url}
\usepackage{amsmath}
\usepackage{empheq}
\usepackage[theorems,skins]{tcolorbox}
\usepackage{wrapfig}

\newtcolorbox{mymathbox}[1][]{colback=white, sharp corners, #1}
\newtcbox{\othermathbox}[1][]{nobeforeafter, math upper, tcbox raise base, enhanced, sharp corners, colback=black!10, colframe=red!30!black, drop fuzzy shadow, left=1em, top=0.5em, right=2em, bottom=0.5em}

\newcommand{\beq}{\bea}  
\newcommand{\eeq}{\eea}  
\newcommand{\bea}{\begin{eqnarray}}  
\newcommand{\eea}{\end{eqnarray}}

\newcommand{\fet}[1]{\mathbf{#1}}

\newcommand{\jr}[1]{{\textcolor{Turquoise}{#1}}}

\begin{document}


\title{Defect self-propulsion in active nematic films with spatially-varying activity}

\author{Jonas R\o nning$^1$, M.~Cristina Marchetti $^2$ and Luiza Angheluta$^1$}

\address{$^1$Njord Centre, Department of Physics, University of Oslo, P. O. Box 1048, 0316 Oslo, Norway\\
$^2$Department of Physics \& Biomolecular Science and Engineering Program, University of California Santa Barbara, Santa Barbara, CA 93106, USA\\
}

\subject{soft matter, biophysics, fluid mechanics}


\corres{Jonas Rønning\\
\email{jonasron@uio.no}}

\begin{abstract}
We study the dynamics of topological defects in active nematic films with spatially-varying activity and consider two setups: i) a constant activity gradient, and ii) a sharp jump in activity. A constant gradient of extensile (contractile) activity endows the comet-like $+1/2$ defect with a finite vorticity that drives the defect to align its nose in the direction of decreasing (increasing) gradient.  A constant gradient does not, however, affect the known self-propulsion of the $+1/2$ defect and has no effect on the $-1/2$ that remains a non-motile particle. A sharp jump in activity acts like a wall that traps the defects, affecting  the translational and rotational motion of both charges.
The $+1/2$ defect slows down as it approaches the interface and the net vorticity tends to reorient the defect polarization so that it becomes perpendicular to the interface. The $-1/2$ defect acquires a self-propulsion towards the activity interface, while the vorticity-induced active torque tends to align the defect to a preferred orientation.  This effective attraction of the negative defects to the wall is consistent with the observation of an accumulation of negative topological charge at both active/passive interfaces and physical boundaries.
\end{abstract}
\maketitle

\section{Introduction} \label{sec:intro}
Active nematics are collections of elongated apolar particles that consume energy from their surroundings to generate dipolar forces that  drive self-sustained  flows~\cite{Doostmohammadi2018}. Much progress in understanding the rich dynamics of these active liquid crystals has been achieved through a minimal hydrodynamic theory that couples orientational order and flow and captures the behavior of biological systems from subcellular to multicellular scales~\cite{balasubramaniam2022active}. Within the biological realm, the active nematic paradigm describes mixtures of cytoskeletal filaments and motor proteins ~\cite{sanchez2012spontaneous,guillamat2017taming,needleman2017active,kumar2018tunable}, bacterial suspensions ~\cite{Doostmohammadi2018, marchetti2013hydrodynamics} and confluent cell monolayers ~\cite{saw2018biological,mueller2019emergence}. 

What distinguishes the hydrodynamics of  active nematics from that of their passive counterparts is the presence of an active stress generated by active processes, which sets up spontaneous spatio-temporally chaotic flows~\cite{marchetti2013hydrodynamics}. The active stress is given by $\sigma^a_{ij}= \alpha Q_{ij}$ with $\mathbf{Q}$ the nematic order parameter  and $\alpha$  a scalar activity parameter that encapsulates the biochemical processes that generate active forces~\cite{simha2002hydrodynamic,juelicher2007active,giomi2013,lemma2019statistical}. It can have either sign: $\alpha>0$ corresponds to a system of "pullers" generating contractile stresses on their surroundings, whereas $\alpha<0$ reflects a system of "pushers" and their induced extensile active stresses. With increasing activity,  active flows are induced spontaneously and create large distortions of the nematic order, including the formation of pairs of topological defects that sustain active turbulence \cite{sanchez2012spontaneous,alert2022active}.

\begin{figure}[t]
    \centering
    \includegraphics[width=0.45\textwidth]{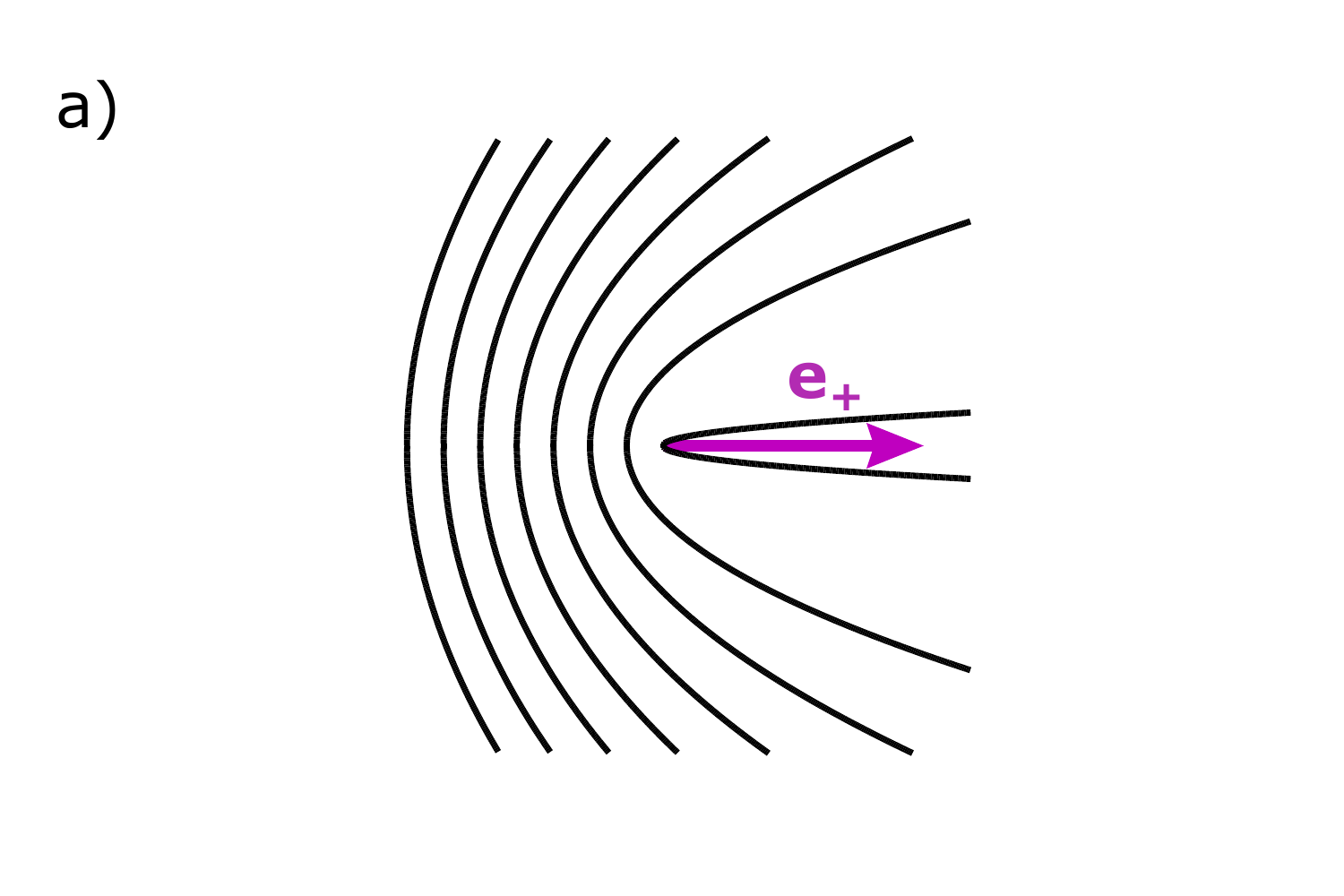}
    \includegraphics[width=0.45\textwidth]{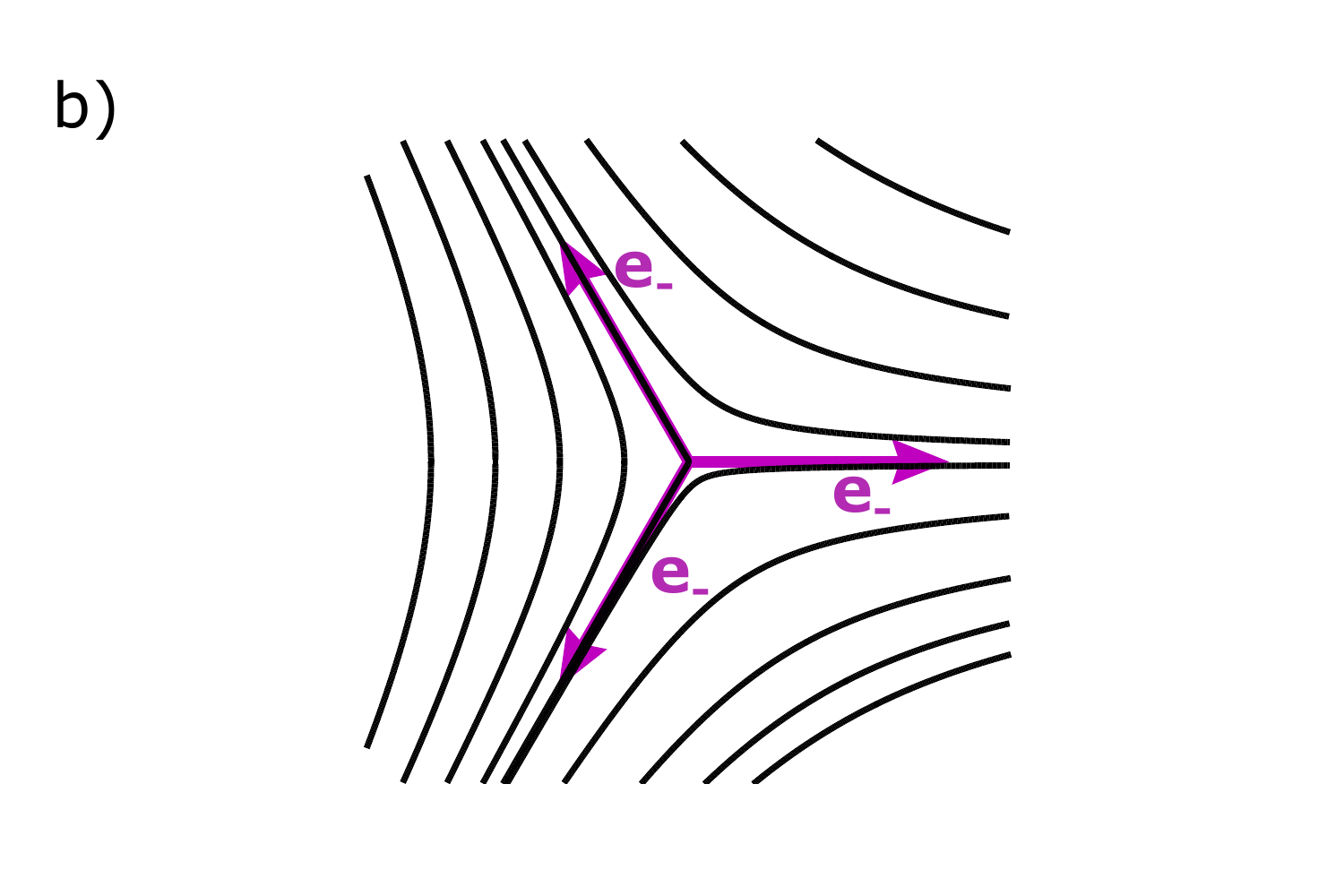}
    \caption{Illustration of (a) the $+1/2$ defect and (b) the $-1/2$ defect and their corresponding polarizations for $\theta_0 = 0$. The negative defect has three equivalent polarizations. } 
    \label{fig:defect_sketch}
\end{figure}

The lowest energy topological defects in active nematic films have half-integer charge, corresponding to the comet-shaped $+1/2$ with polarity $\mathbf{e}_+$ defined by a head-tail arrow, and the $-1/2$ which has three-fold symmetry (see Fig.~\ref{fig:defect_sketch}). These defects disrupt the nematic order locally and induce long-range  distortions in the orientation field, generating  active stresses, which in turn lead to spontaneous active flows surrounding the defects~\cite{giomi2013,pismen2013,ronning2022flow}. There is a net active flow through the core of the polar $+1/2$ defect which makes it intrinsically motile and is referred to as the defect self-propulsion. For an isolated $+1/2$ defect, the self-propulsion velocity aligns with the polarity vector and, depending on the contractile/extensile properties of the active nematic, the defect moves in/opposite to the direction of its polarization. The $-1/2$ defect does not create any net flow at the defect position and, thus, is not self-propelled in systems with uniform activity. The motion of defects in the presence of spatially inhomogeneous activity is far less understood and explored~\cite{shankar2019hydrodynamics}.  

There are several approaches to realize experimentally systems with spatially-dependent activity. In Ref.~\cite{thijssen2021submersed},  a varying substrate topography is used   to control the frictional damping in a film of a microtubule-kinesin suspension. This results in spatial variations of the concentration of active agents, thus indirectly in the local activity. In Ref.~\cite{zhang2021spatiotemporal} a similar effect is achieved by manipulating light-sensitive myosin motors that activate the microtubules. Both studies find that the  $-1/2$ defects localize near the interface separating the region of higher activity from that of lower activity. In Ref.~\cite{zhang2021spatiotemporal}, it was reported that the $+1/2$ defects are deflected by the active/passive interface.
Analytical work based on a hydrodynamic theory of the defect gas has predicted that a passive/active interface can be used to separate positive and negative topological charge~\cite{shankar2019hydrodynamics}.
Numerical studies of how the defect dynamics is affected by the spatially-dependent activity show that the polarity of the $+1/2$ defect tends to align parallel to the activity gradient~ \cite{tang2021alignment,mozaffari2021defect,zhang2021spatiotemporal,zhang2022logic,ruske2022activity}, and that the confinement and motion of defects can be manipulated by varying the steepness of the activity gradients~\cite{LSthesis,scharrer2022spatial}. A recent numerical study also shows that the formation of defect dipoles can be controlled by imprinting special geometries into the activity profile~\cite{zhang2022logic}.

In this paper, we provide a theoretical study of how spatially-varying activity affects the self-propulsion and reorientation of isolated topological defects. We consider the representative basic setups where the activity gradient is either constant or a Dirac delta function (sharp interface). For constant activity gradients, the $+1/2$ defect rotates due to a vorticity-induced active torque acting on the defect polarization until the defect aligns parallel to the activity gradient, and moves in the direction of lower magnitude of activity.
Thus the defect slows down. We show analytically that the vorticity at the $+1/2$ defect core is proportional to the hydrodynamic dissipation length $\ell_d=\sqrt{\eta/\Gamma}$, which measures the strength of viscous dissipation $\eta$ relative to friction $\Gamma$.
Numerical simulations of the active flow generated in a disk of radius $R$ show that the vorticity depends  on the system size for small $R$, and crosses over to the analytically-predicted value for large systems. The vorticity field induced by an activity gradient parallel to the $+1/2$ defect's polarization has a quadruple structure with regions of alternating vorticity. This is confirmed by numerical simulations for a disk geometry where four vortices are formed around the $+1/2$ defect. 
In contrast, the vorticity induced by constant activity gradients at a $-1/2$ defect has an $8$-fold symmetry that leads to eight vortices with alternating circulation in a finite domain.  We also calculate  the net translational self-propulsion and reorientation that both  $\pm 1/2$ defects acquire near a sharp active/passive interface.
The $+1/2$ defect slows down as it moves towards the interface, and the vorticity-induced torque tends to re-orient it such that its polarization becomes normal to the interface regardless of the sign of activity. The $-1/2$ defect also acquires a preferred orientation at the interface, and those that approach the interface with this stable orientation are then attracted by it. 

The structure of the paper is as follows. We start in Section \ref{Sec:hydro} by introducing the minimal hydrodynamic model active nematic films on a substrate. In Section~\ref{sec:Constant gradient}, we derive and discuss the self-propulsion and spontaneous rotation of $+1/2$ defects in the presence of a  constant activity gradient. Section~\ref{sec:wall} focuses on the analytical derivation of the self-propulsion and vorticity of $\pm 1/2$ defects close to a sharp active/passive interface. Summary and concluding remarks are presented in Section \ref{sec:conclusion}.

\section{Hydrodynamics of active nematics with spatially-varying activity}\label{Sec:hydro}
We consider the familiar hydrodynamic model of a 2D active nematics that couples the flow velocity $\fet u(\fet r)$ to  the nematic order parameter $Q_{ij} = S (\hat n_i \hat n_j-\frac{1}{2} \delta_{ij})$, where $S$ quantifies the degree of order and $\mathbf{\hat n}(\mathbf{r})=\left(\cos\theta(\mathbf{r}),\sin\theta(\mathbf{r})\right)$ is the orientational director field with head-tail symmetry. In the simplest formulation, the $\textbf{Q}$-tensor is a minimizer of the de Gennes-Landau free energy ~\cite{marchetti2013hydrodynamics}
 \begin{align}
\mathcal F = \int d\fet r \left[\frac{K}{2}|\partial_i Q_{jk}|^2+\frac{g}{4} \textrm{Tr}\left(1-\textrm{Tr}\left(\mathbf{Q}^2\right) \right)^2\right]\;,
\label{eq:F}
 \end{align}
with isotropic elastic constant $K>0$ and $g$ the strength of the local ordering potential. The uniform nematic ordered state  corresponds to $S_0=1$.
The  flow field  satisfies a  Stokes equation that balances forces on a fluid element, given by~\cite{marchetti2013hydrodynamics}
\begin{align}
    \left(\Gamma-\eta \nabla^2 \right) \fet u =  \bm\nabla\cdot[\alpha(\fet r)  \mathbf{Q}(\fet r)] -\bm\nabla p(\fet r), \qquad \bm\nabla\cdot \fet u =0\;,
    \label{eq:Stockes_units}
\end{align}
where  $\Gamma$ is a friction coefficient per unit area,  $\eta$ is the shear viscosity, and $\alpha$ is the activity coefficient, with dimensions of stress. For simplicity, we neglect additional passive elastic stresses, which are of higher order in the gradients of $\mathbf{Q}$ compared to the active stress and a more important contributions for nematic textures with many defects. Instead we focus on the active flows generated by an isolated $\pm 1/2$ in the presence of non-homogeneous activity $\alpha(\fet r)$.  

We rescale the Stokes equation in units of the nematic relaxation time $\tau =\gamma/g $ (where $\gamma$ is the nematic  rotational friction)~\cite{angheluta2021role, ronning2022flow} and the coherence length $\xi=\sqrt{K/g}$. Different dynamical regimes are then controlled by one dimensionless number $\zeta = \ell_d/\xi$, where $\ell_d =\sqrt{\eta/\Gamma}$, and the rescaled activity   $\alpha(\mathbf r) \rightarrow \alpha(\mathbf r)\gamma/(\Gamma K)$. The dimensionless form of the Stokes equation reads as

\begin{equation}
    \left(1- \zeta^{2}\nabla^2 \right)\fet u =   \fet F_{\pm} -\bm\nabla p\;, \qquad \bm\nabla \cdot \fet u=0\;,
    \label{eq:Stokes}
\end{equation}
where the active force field induced by an isolated $\pm 1/2$ defect is given by
\begin{equation}
    \fet F_{\pm} =  \mathbf{Q}(\fet r)\cdot \bm\nabla\alpha(\fet r)  +  \alpha(\fet r)  \bm\nabla \cdot \mathbf{Q}(\fet r) =\mathbf F_{\pm}^I + \mathbf F_{\pm}^B\;. \label{eq:active_force}
\end{equation}
The first contribution is an interfacial force $\mathbf F_{\pm}^I$ originating from activity gradients. The second term is a bulk force $\mathbf F_{\pm}^B$ due to nematic distortions. The defect self-propulsion velocity $\fet v_{\pm}$ is defined as the net active flow through the defect core, and thus can be computed from the active flow velocity $\mathbf u$ obtained from the solution of Eq.~(\ref{eq:Stokes}) evaluated at the origin \cite{angheluta2021role}. The solution of Eq.~(\ref{eq:Stokes}) is given by
\begin{equation}
\fet v_{\pm}  = \frac{1}{2\pi\zeta^2} \int d\mathbf r K_0\left(\frac{r}{\zeta}\right) \left[\fet F_{\pm}(\mathbf r) -\bm\nabla p(\mathbf r)\right] = \mathbf v_{\pm}^I+\mathbf v_{\pm}^B,
\label{eq:defect_velocity}
\end{equation}
where $K_0(r)$ is the zeroth order Bessel function which is the Green's function of Eq.~(\ref{eq:Stokes}) without the incompressibility constraint, and  
the pressure field is the solution of the corresponding Poisson's equation
\begin{equation}
    \nabla^2 p =  \bm\nabla\cdot\fet F_{\pm}(\fet r)\;. 
    \label{eq:pressure}
\end{equation}
The net vorticity at the defect core is also obtained from measuring the vorticity of the flow field induced by the defect distortion, given by $\omega = \partial_x u_y -\partial_y u_x = -\nabla^\perp \cdot \mathbf u$. Using Eq.~(\ref{eq:defect_velocity}) and evaluating it at the defect position $\fet r_0=0$, we obtain an expression for the defect vorticity 
\begin{equation}
\mathbf \omega_{\pm}  = -\frac{1}{2\pi\zeta^2} \int d\mathbf r K_0\left(\frac{r}{\zeta}\right) \bm\nabla^\perp \cdot \fet F_{\pm}(\mathbf r)= \mathbf \omega_{\pm}^I+\mathbf \omega_{\pm}^B\;.
\label{eq:defect_vorticity}
\end{equation}
Both  velocity and  vorticity are written as sums of interfacial and bulk contributions which depend on the defect polarization $\fet e_{\pm}$ and are computed analytically in the next sections. 

For isolated $\pm 1/2$ point-like defects, we can parameterise the $\mathbf Q$-tensor order-parameter in the quasistatic phase approximation as \cite{angheluta2021role}
\begin{equation}
    Q^\pm_{xx}(\mathbf r) = \cos(\pm\phi(\mathbf r) + 2\theta_0)\;, \qquad
    Q^\pm_{xy}(\mathbf r) = \sin(\pm\phi(\mathbf r) + 2\theta_0)\;, \label{eq:Q}
\end{equation}
where $\phi(\mathbf r) = \arctan(y/x)$ is the singular part of the nematic orientation due to a $\pm 1/2$ defect located at the origin, and $\theta_0$ is the slowly-varying part of the background orientation of the nematic director. The $+1/2$ defect has a well-defined polarization which is determined by the background nematic orientation $\theta_0$ as
\beq
\fet e_{+} =\left(\frac{ \bm\nabla \cdot \mathbf{Q}}{|\bm\nabla \cdot \mathbf{Q}|}\right)_{\mathbf r = 0} = [\cos(2\theta_0), \sin(2\theta_0)].
\label{eq:positive_polarization}
\eeq
For the $-1/2$ defect, we can also introduce a polarization vector determined by $\theta_0$ and aligning with one of the principal axes of the three-fold symmetry \cite{angheluta2021role}
\beq
\fet e_{-} = [\cos(2\theta_0/3), \sin(2\theta_0/3)]. \label{eq:negative_polarization}
\eeq
Both nematic defects and their respective polarizations are illustrated in Fig.~\ref{fig:defect_sketch}.

It can be shown that a net vorticity at the defect core induces an active torque that tends to rotate the defect polarization. This follows straightforwardly from taking the time derivative of the polarization in Eqs.~ (\ref{eq:positive_polarization}) and (\ref{eq:negative_polarization}), and using the evolution of the $\mathbf Q$-tensor~\cite{shankar2019hydrodynamics,angheluta2021role} to account for the change in the background nematic field $\theta_0$ due to vorticity as   $\partial_t\theta_0\approx \omega/2$. Thus, the evolution of the defect polarization controlled by vorticity is
\begin{equation}
    \mathbf{\dot e}_\pm \approx -3^{-1/2 +q}\omega_{\pm}\mathbf e_\pm^\perp, \label{eq:active_torque}
\end{equation}
where the defect charge is $q=\pm 1/2$ and $\fet e^\perp = [e_y,-e_x]$ represents the $90^0$ clockwise rotation of the polarization vector. For motile defects, there are additional torques due to defect interactions, the elastic stiffness $K$ or the coupling to the flow alignment~\cite{shankar2019hydrodynamics, angheluta2021role}. Here, we focus on the active torque induced by a non-zero vorticity which emerges from spatially varying activity  alone. In the subsequent sections, we investigate how this active torque reorients the defect polarization relative to activity gradients for two setups: i) a constant activity gradient, and ii) an interface with a sharp jump in activity.

\section{Constant activity gradient} \label{sec:Constant gradient}

We first study the kinematics of an isolated defect in a region where the activity gradient is locally constant. Without loss of generality, we consider an activity gradient in the $x$-direction such that the activity has the linear profile $\alpha(\mathbf r) = \alpha_0 + \alpha_g x$. The defect orientation is arbitrary and controlled by the background nematic orientation $\theta_0$. We demonstrate that a constant gradient $\alpha_g$ does not modify the defect  self-propulsion velocity as compared to what was obtained for uniform bulk activity $\alpha_0$. An activity gradient across the texture of a  $+1/2$ defect generates, however, a flow that may yield a finite vorticity at the defect core, which tends to align the defect polarization according to Eq.~(\ref{eq:active_torque}) in the direction of the gradient. The $-1/2$ defect remains stationary both in its motion and orientation. 

\begin{figure}[t]
    \centering
    \includegraphics[width = 0.45\textwidth]{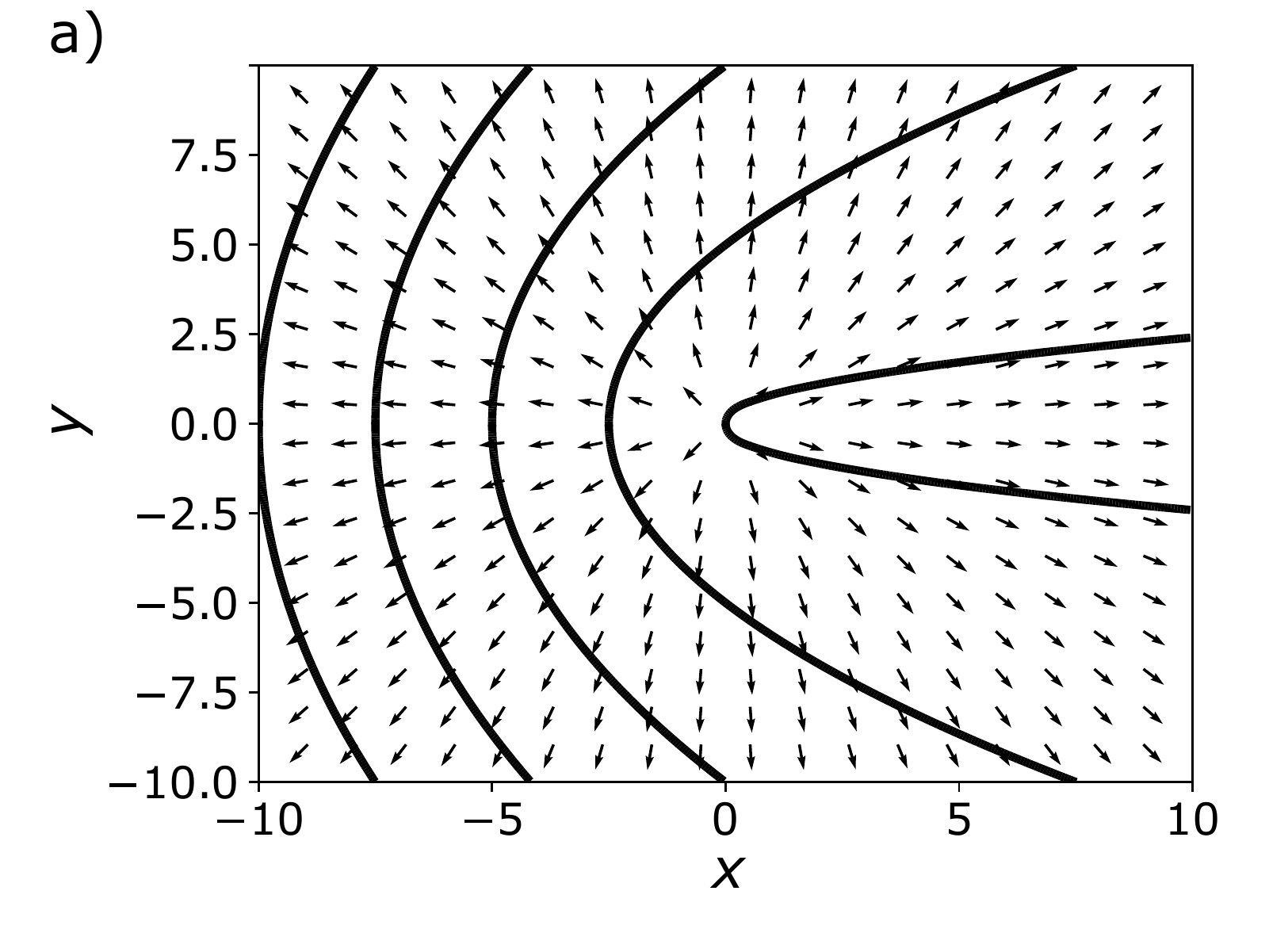}
    \includegraphics[width = 0.45\textwidth]{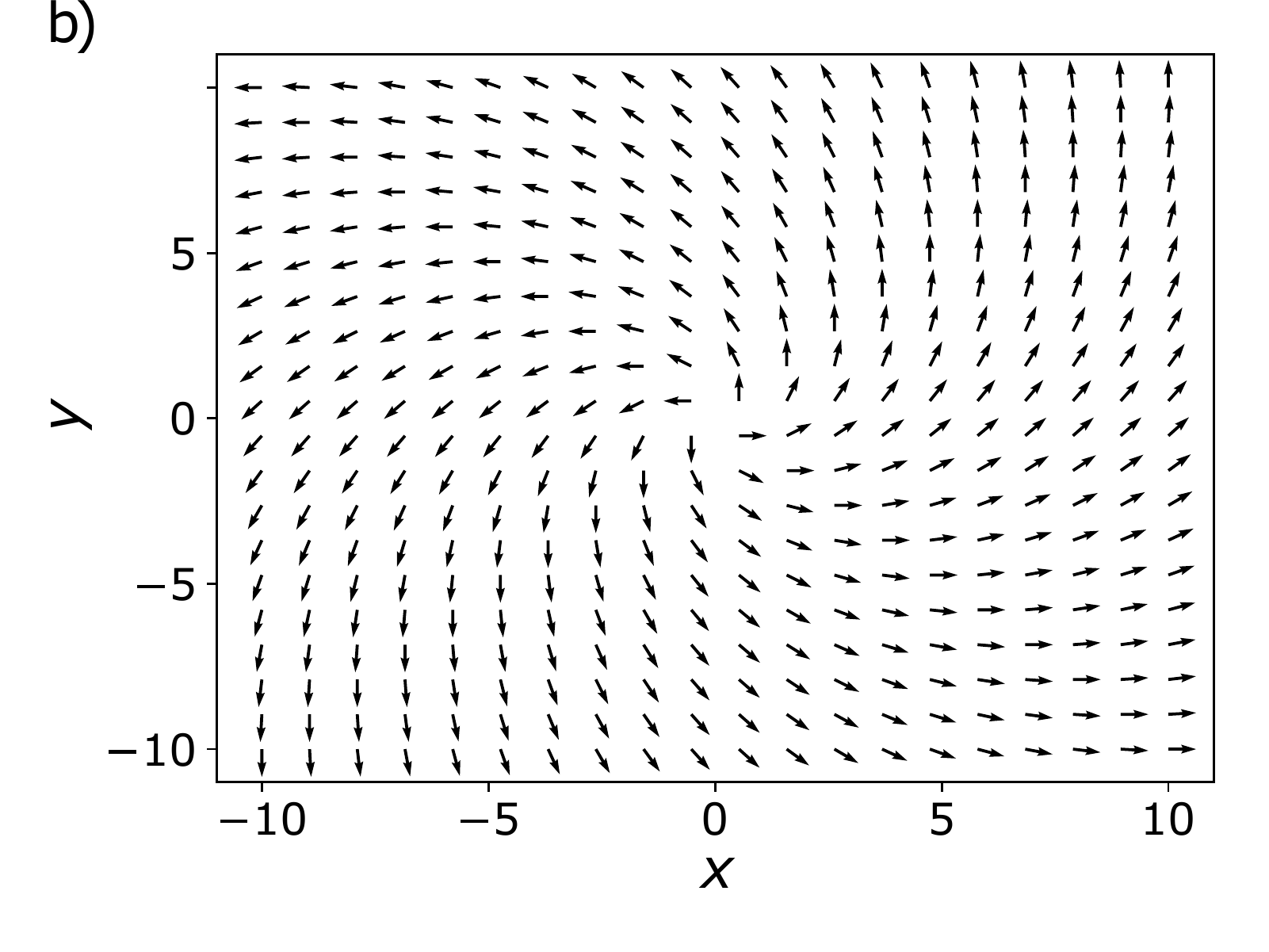}
    \includegraphics[width = 0.45\textwidth]{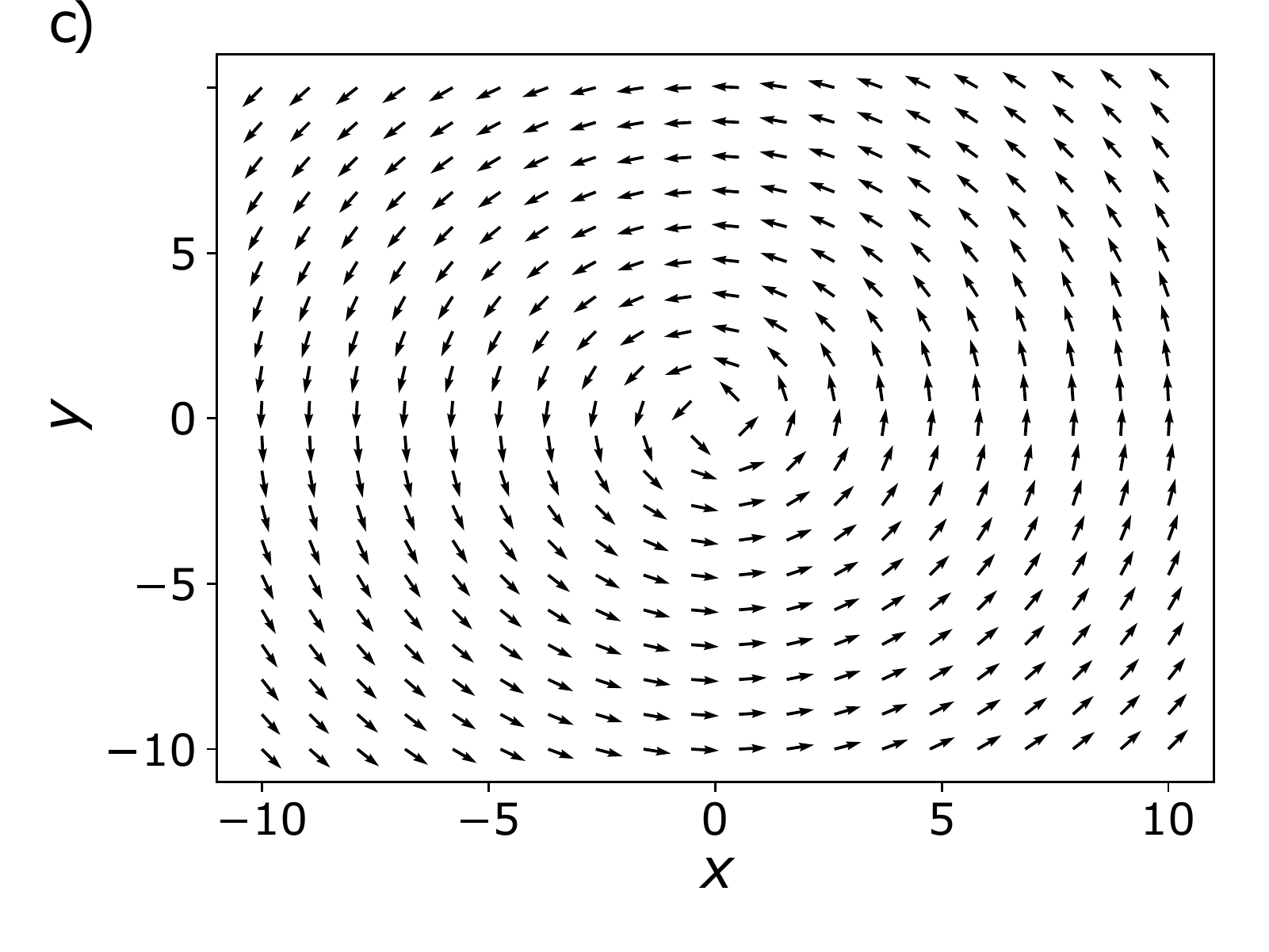}
    \caption{Interfacial active force field from Eq.~(\ref{eq:grad_force}) induced by a $+1/2$ defect with  a) $\theta_0 = 0$, b) $\theta_0 =\pi/8$ and c) $\theta_0 = \pi/4$. Note that cases b) and c) lead to rotation of the defect together with the nematic field until the defect polarization aligns with the direction of the activity gradient. The dark solid lines in a) show the nematic field around the $+1/2$ defect oriented in the $x$ direction.  }
    \label{fig:gradient_forces} 
\end{figure}

\subsection{$+1/2$ defect}
The interfacial active force given in Eq.~(\ref{eq:active_force}) arising from  a constant activity gradient  $\alpha_g$ is
\begin{equation}
    \mathbf F_{+}^I (\fet r)=\alpha_g \left[ \cos(2\theta_0)\mathbf{\hat r}  -  \sin (2\theta_0)\mathbf{\hat r}^\perp  \right]\;,
    \label{eq:grad_force}
\end{equation}
where $\fet{\hat r}$ is the radial unit vector and $\mathbf{\hat r}^\perp = (\hat y,-\hat x)$. This expression corresponding to the Helmholtz decomposition of  $\mathbf F_{+}^I$ into a curl-free part ($\sim\fet{\hat r}$) and a divergence-free part ($\sim\fet{\hat r}^\perp$). These two contributions are plotted in
Fig.~(\ref{fig:gradient_forces}). The divergence-free part gives a net vorticity at the defect core which tends to rotate its polarization until it aligns with the activity gradient. This is most easily demonstrated in the friction-dominated limit where the active flow velocity is  $\Gamma \mathbf u = \mathbf F_{+}^I-\bm\nabla p$. The incompressibility constraint thereby removes the curl-free contribution through the contribution of the interfacial pressure which is radially symmetric and given by
\begin{equation}
p_+^I (\mathbf r)= \alpha_g \cos(2\theta_0)  (r - L),
\end{equation}
making the interface flow purely rotational. Here the constant $L$ is a length comparable with the system size which controls the divergent terms. More generally, to incorporate viscous dissipation we need to evaluate the integral expression for the defect velocity given in Eq.~(\ref{eq:defect_velocity}).  In an infinite system, the symmetry of the integrand leads to no contribution to the defect speed from the interfacial active force, thus $\mathbf v^I_+ =0$.  This contribution may become finite in non-radially symmetric bounded domains.

The contribution from the bulk active force in  Eq.~(\ref{eq:active_force}) reduces to
\begin{equation}
    \mathbf F_+^B (\fet r)= \mathbf F_+^0(\fet r) +\alpha_g x \bm\nabla \cdot \mathbf{Q}_+ = \mathbf F_+^0(\fet r)+ \alpha_g\frac{x}{r} (\cos(2\theta_0)\fet{\hat x} +\sin(2\theta_0) \fet{\hat y} )\;, 
\end{equation}
where $\mathbf{F}_+^0(\fet r)$ is the known active force corresponding to a constant activity $\alpha_0$ which leads to a constant self-propulsion velocity~\cite{angheluta2021role,ronning2022flow}. The contribution due to $\alpha_g$ is anti-symmetric around the defect position, thus its integral over an infinite domain vanishes. Therefore, there is no contribution from activity gradients to the defect self-propulsion. This is not changed when we add the gradient of the bulk pressure which is given as
\begin{equation}
     p_+^B(\mathbf r) = p_+^0 (\mathbf r) + \frac{\alpha_g  \cos(2\theta_0) }{6} \left(  \frac{(x^2 -y^2)}{r} + 3(r-L) \right)-\frac{\alpha_g  \sin(2\theta_0)}{3}   \frac{xy}{r}
\end{equation}
Here $p^0_+$ is the pressure for the constant activity term $\alpha_0$~\cite{ronning2022flow}. 

\begin{figure}[t]
    \centering
      \includegraphics[width = .6\textwidth]{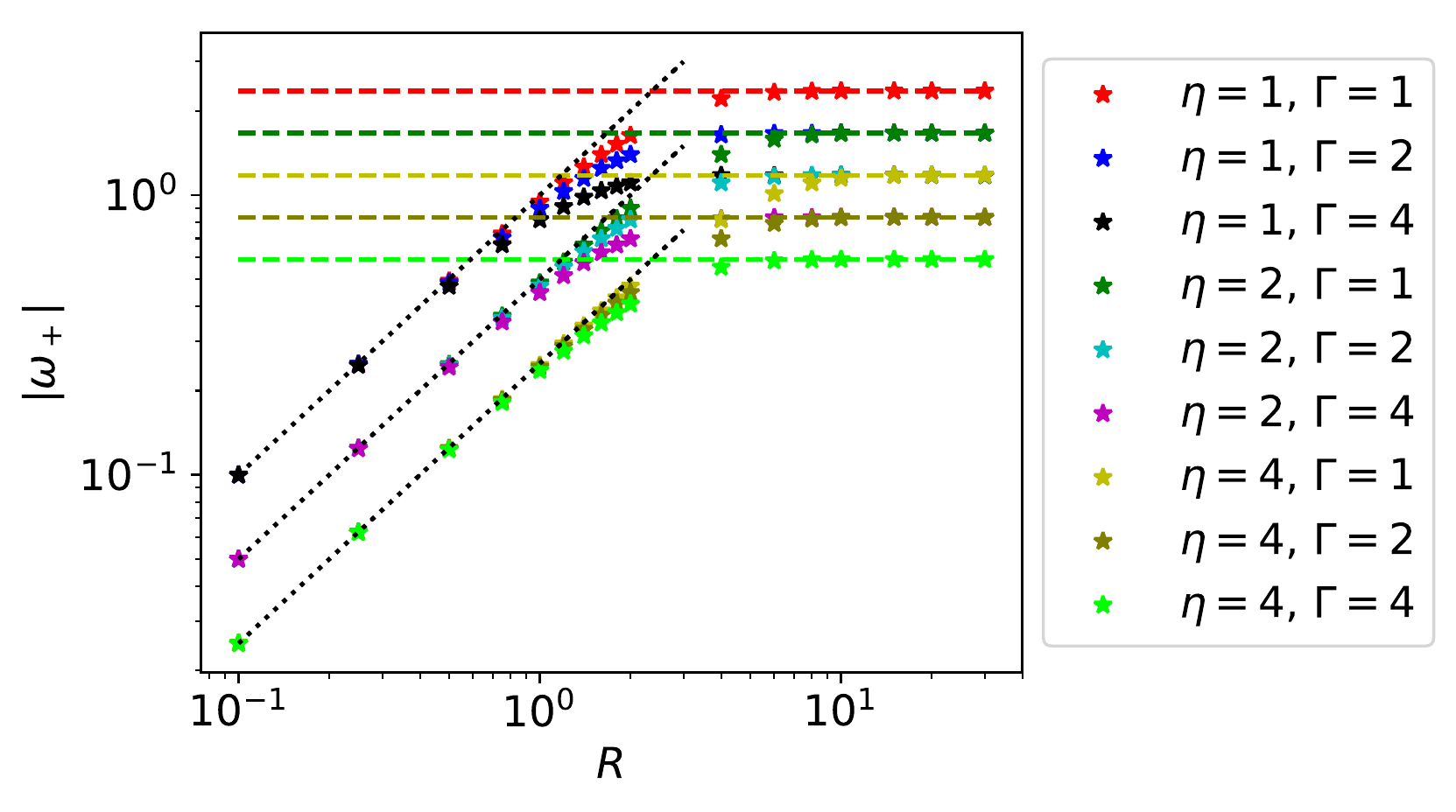}
    \caption{Magnitude of the angular velocity of the $+1/2$ defect  for different values of dissipation parameters $\eta$ and $\Gamma$,  $\alpha_g=1$ and $\theta_0 =-\pi/4$. The dashed horizontal lines are the analytical prediction for an unbounded domain. The dotted lines show the linear scaling with $R$ and with slopes $1/\eta$.}
    \label{fig:vort}
\end{figure}

Activity gradients induce, however, a vortical flow that is finite at the defect core, resulting in an angular velocity of the $+1/2$ defect, given by

\begin{equation}
    \omega_+ =\frac{\alpha_g}{2\pi\zeta^2}\sin(2\theta_0) \int d\mathbf r ~K_0\left(\frac{r}{\zeta}\right) \left(\frac{1}{r}+\frac{x^2}{r^3} \right)\;,
    \label{eq:omega-plus}
\end{equation}
where the first term in the bracket originates from the interfacial active force and the second is due to the bulk force. The integral can be  carried out in polar coordinates, with the result
\begin{equation}
    \omega_+ =\frac{3\pi\alpha_g}{4\zeta} \sin(2\theta_0)\;.
\end{equation}
We can rewrite this equivalently in physical units as
\begin{equation}
    \omega_+ = \frac{3\pi \alpha_g}{4\Gamma\ell_d}\sin(2\theta_0) = \frac{3\pi \alpha_g}{4\sqrt{\Gamma \eta}}\sin(2\theta_0) =\frac{3\pi \alpha_g}{4\eta}\ell_d\sin(2\theta_0), \label{eq:vorticity_positive}
\end{equation}
to highlight that the defect angular velocity scales linearly with the hydrodynamic dissipation length $l_d$, similar to the self-propulsion speed of a defect in a constant activity \cite{ronning2022flow}. The effect of this vorticity is to align the polarization so that it is pointing opposite to the activity gradient. This is consistent with recent numerical results, where defects align normal on soft interfaces separating extensile and contractile regions~\cite{ruske2022activity}.  

To test the validity of these analytical predictions for a bounded system, we have solved numerically  the Stokes Eq.~(\ref{eq:Stokes}) in a disk of radius $R$. In Fig.~(\ref{fig:vort}), we show that the defect angular velocity is proportional to $R$ for radii smaller than $l_d$,  and crosses over to the asymptotic value for an infinite system given by Eq.~(\ref{eq:vorticity_positive}) at large $R$. 

\begin{figure}[t]
    \centering
    \includegraphics[width =1.0 \textwidth, trim = 0.0cm 12.0cm 0.0cm 0.0cm, clip = true]{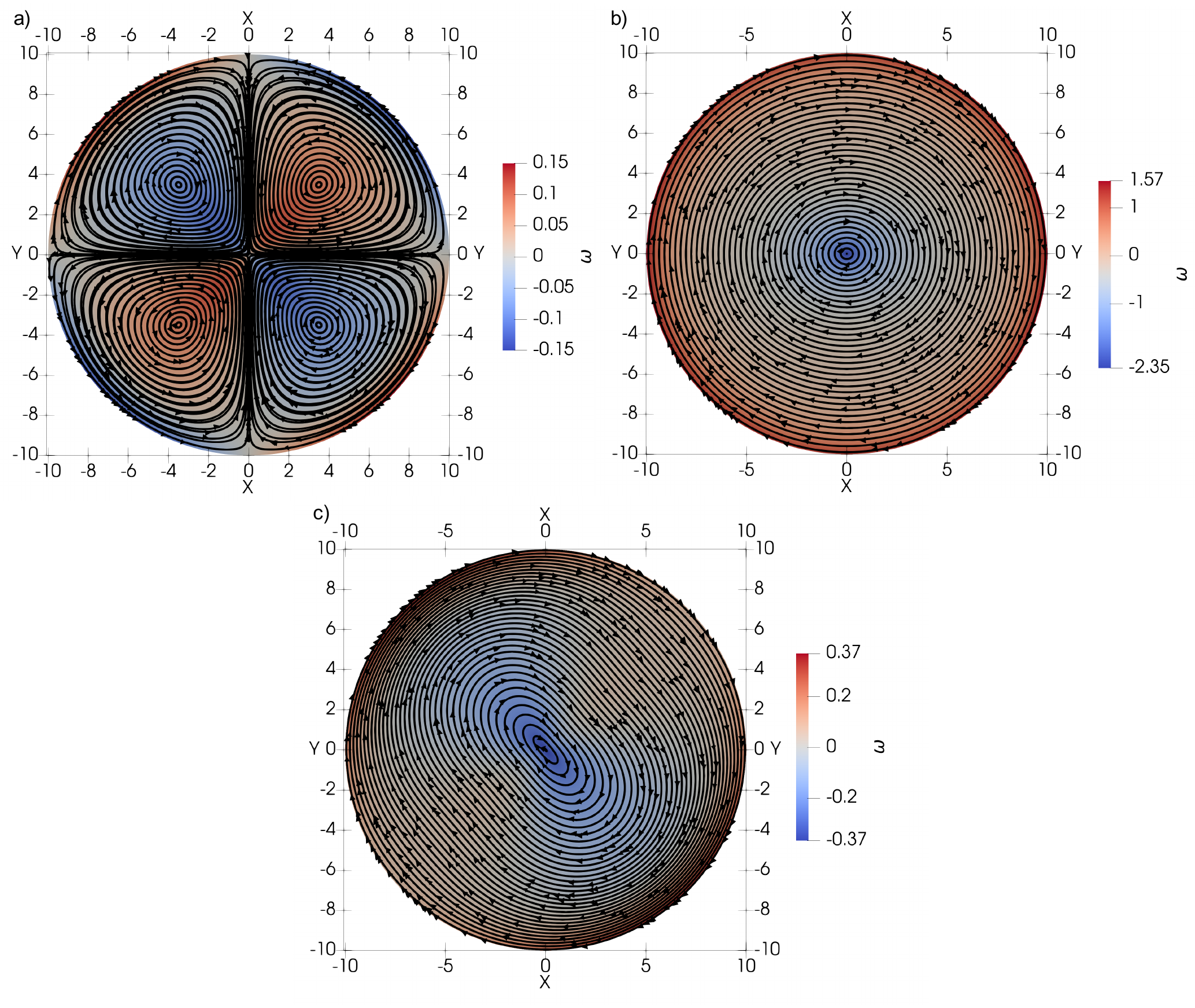}
    \caption{Active flow streamlines induced by a uniform activity gradient along x-direction with $\alpha_0=0$ and by a $+1/2$ defect with orientation (a) $\theta_0 =0$, (b) $\theta_0 = -\pi/4$, and (c) $\theta = -\pi/40$. Since $\alpha_0=0$, the flow velocity vanishes at the defect core. The background colormap represents the vorticity field. When $\theta_0 \neq 0$, the defect acquires a non-zero vorticity at its center as predicted theoretically. The four-fold vortex structure is only visible for small values of $\theta_0$, i.e., when the defect is closely aligned with the direction of the activity gradient. }
    \label{fig:Fem}
\end{figure}

We have also computed the vorticity field for $\alpha_0=0$ and different defect orientations relative to the activity gradient, as shown in  Fig.~(\ref{fig:Fem}). 
The systems are solved using the finite element package FEniCS \cite{alnaes2015fenics,LoggMardalEtAl2012}.
When the defect polarization is parallel to the activity gradient ($\theta_0=0$), we observe a quadruple structure of the vortical flow. This is consistent with the analytical prediction in the friction-dominated limit, where the vorticity field away from the defect is determined by the activity gradient $\alpha_g$ as (for $\alpha_0=0$)
\begin{equation}
    \omega_+(r,\phi) = \frac{\alpha_g \sin(2\phi)}{2\Gamma r} \cos(2\theta_0) + \frac{\alpha_g}{\Gamma r} \sin(2\theta_0)\left(1 +\cos^2(\phi)\right)\;,
\end{equation}
where $r$ and $\phi$ are the polar coordinates centered at the defect position. In contrast, when the defect polarization is normal to the activity gradient ($\theta_0=\pi/2$), we obtain a single vortex centered at the core of the defect.

\subsection{$-1/2$ defect}

\begin{figure}
    \centering
    \includegraphics[width =\textwidth, trim = 0.0cm 21.0cm 0.0cm 0.0cm, clip = true]{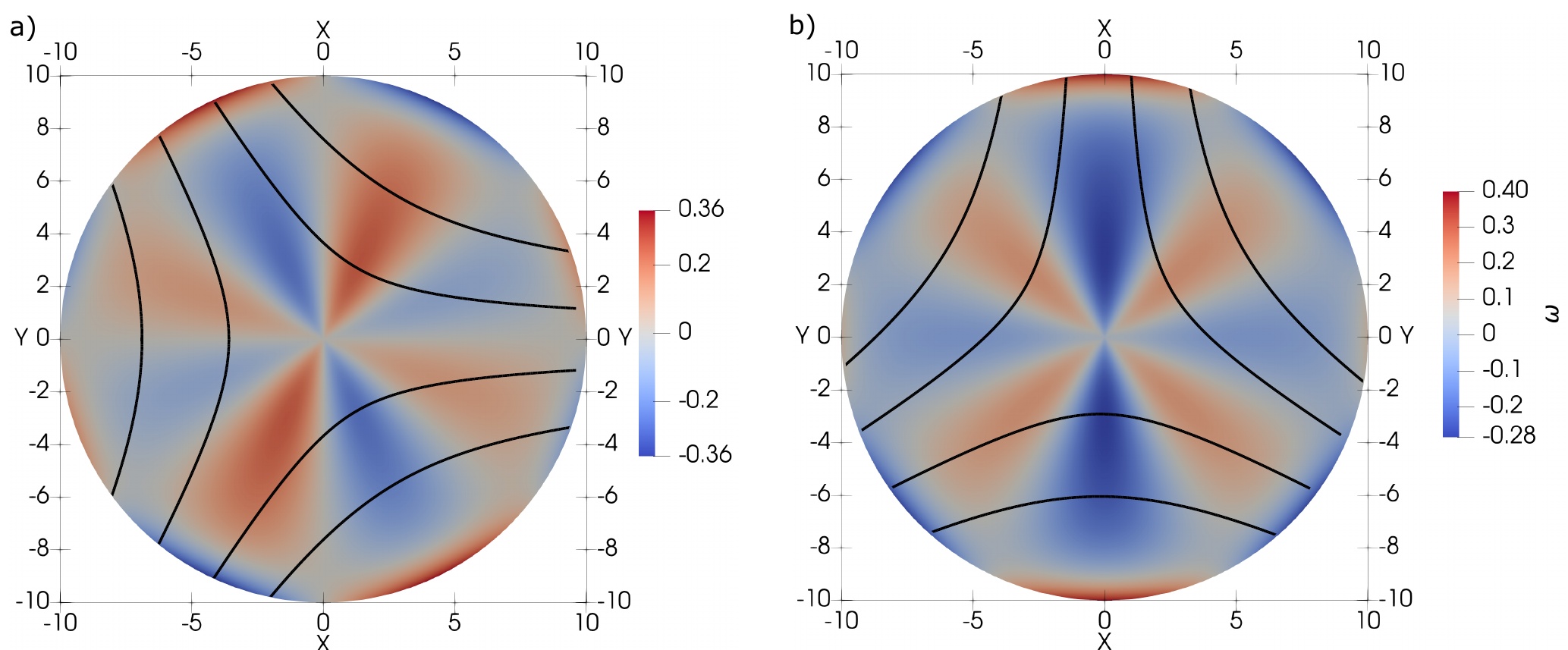}
    \caption{The vorticity field induced by an uniform activity gradient along the x-direction for $\alpha_0=0$ and a $-1/2$ defect with orientation (a) $\theta_0 =0$  and (b) $\theta_0 = -\pi/4$. The black lines shows the director field.  }
   \label{fig:fem_neg}
\end{figure}
A similar analytical calculation can be carried out for the $-1/2$ defect using the parameterization of the $\mathbf Q$-tensor in Eq.~(\ref{eq:Q}). The interfacial and bulk components of the active force field are obtained from Eq.~(\ref{eq:active_force})  as 
\bea
    \mathbf F_{-}^I(\fet r) &=& \frac{\alpha_g}{r} \left[(x \fet{\hat x} -y\fet{\hat y})\cos(2\theta_0) +(y\fet{\hat x} +x\fet{\hat y}) \sin(2\theta_0)  \right], \\
      \mathbf F^B_-(\fet r) &=& \frac{\alpha_g x}{r^3}\left[ (y^2 -x^2)(\cos(2\theta_0)\fet{\hat x}+\sin(2\theta_0)\fet{\hat y}) + 2xy( \cos(2\theta_0) \fet{\hat y} - \sin(2\theta_0)\fet{\hat x)} \right]\;.
\eea
From symmetry considerations these forces as well as their curl  vanish upon integration. This  implies that a constant activity gradient alone does not induce any self-propulsion of the $-1/2$ defect nor a rotation of its orientation. Including the pressure contributions does not alter this effect. 

In the friction-dominated limit and for $\alpha_0=0$, we can evaluate the vorticity field, and show  that a constant activity gradient $\alpha_g$ induces eight counter rotating vortices, with a vortical flow given by
\begin{equation}
    \omega_-(r,\phi) = -\frac{\alpha_g \cos(2\theta_0)}{2\Gamma r}(3\sin(4\phi)-\sin(2\phi)) +\frac{\alpha_g \sin(2\theta_0)}{2\Gamma r}\left(3\cos(4\phi) - \cos(2\phi)\right)\;. 
\end{equation}
The same structure is observed in bounded domains where the vorticity form vortices of alternating circulation, as shown in Fig.~(\ref{fig:fem_neg}) for a disk geometry.

\section{Activity jump at an interface}
\label{sec:wall}

\begin{figure}[t]
    \centering
   \includegraphics[width = 0.4\textwidth]{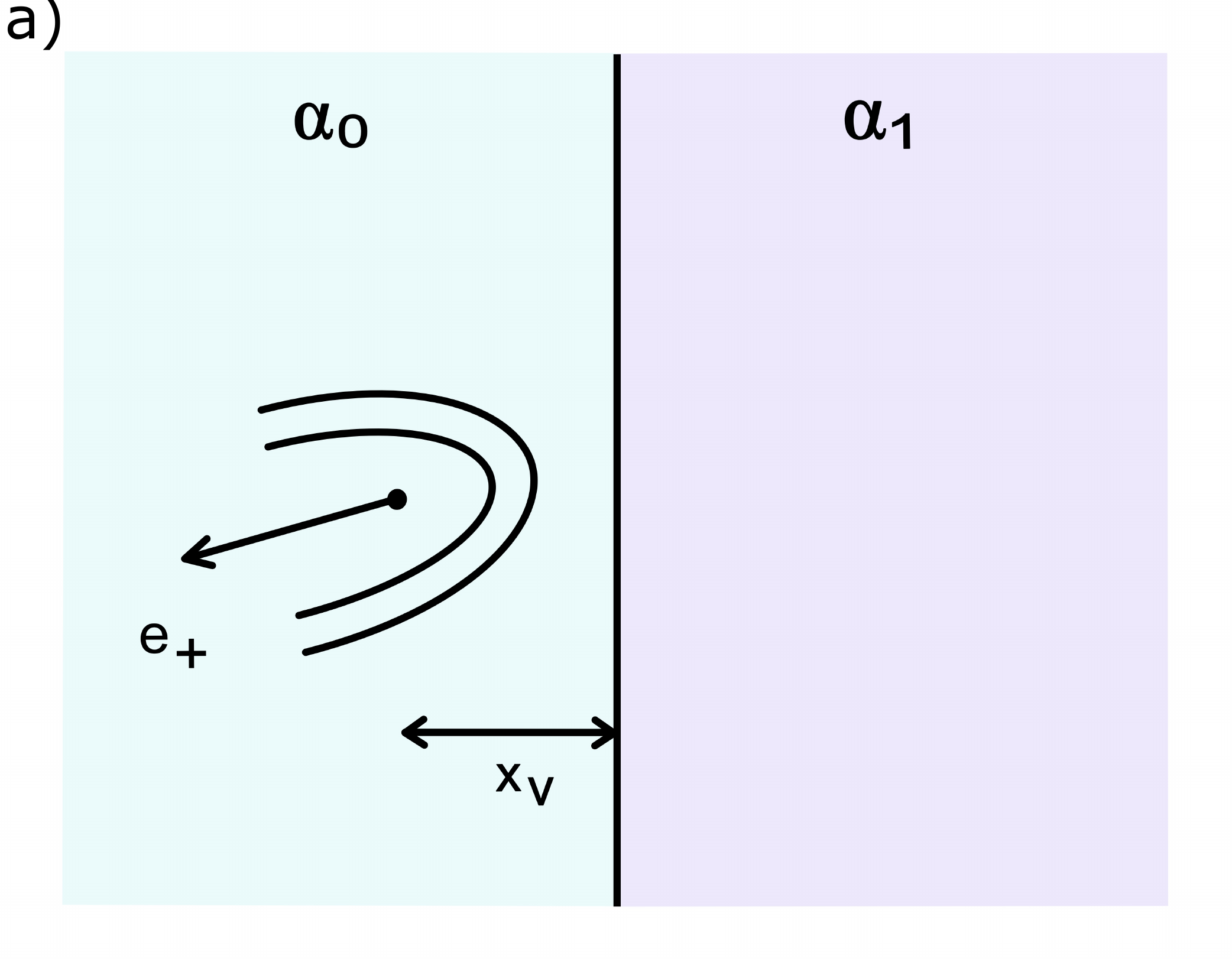}
   \includegraphics[width = 0.4\textwidth]{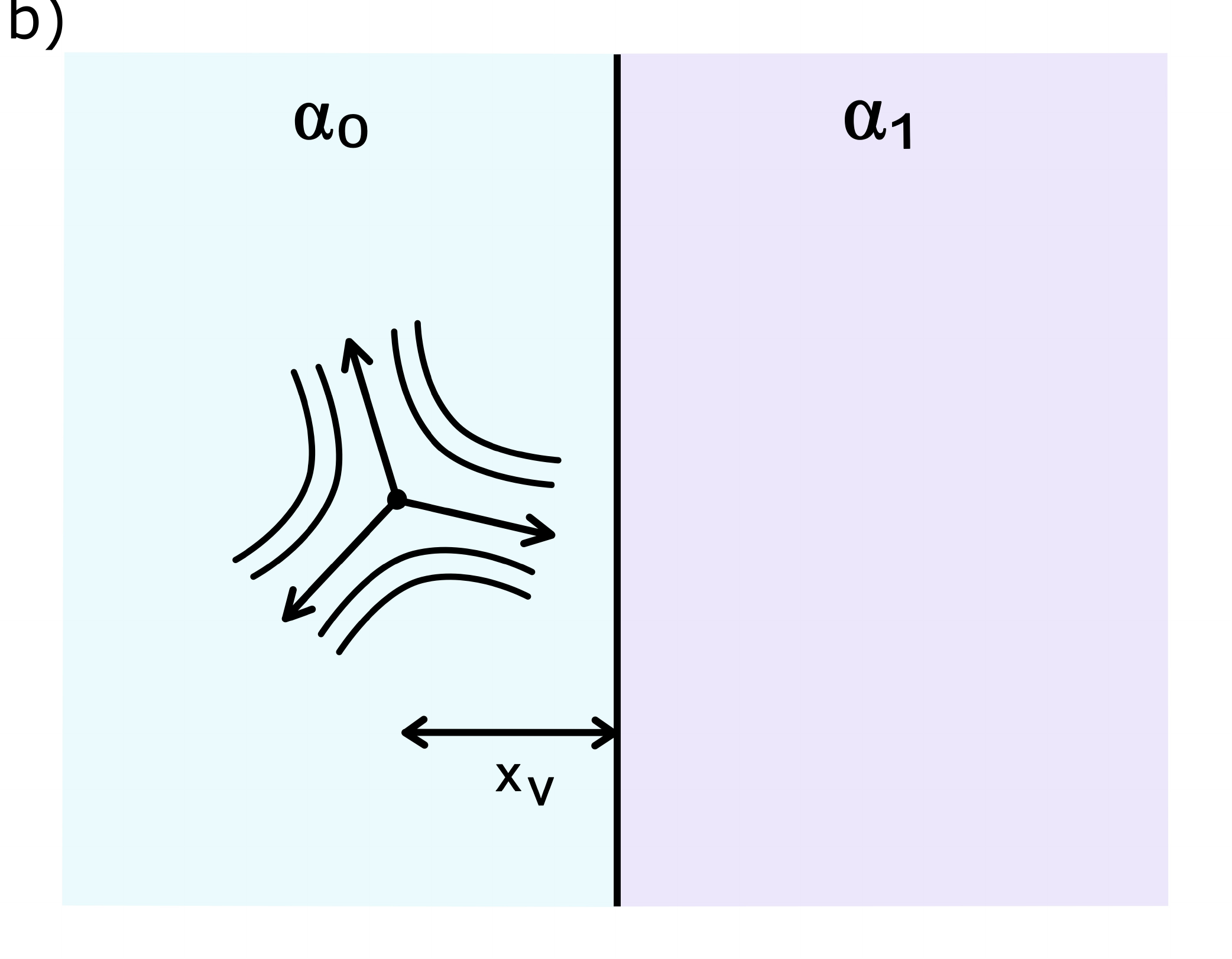}
   \caption{Setup of (a) $+1/2$ defect and (b) $-1/2$ defect at a sharp interface separating a region with higher activity $\alpha_0$ from that  with lower activity $\alpha_1$.}
   \label{fig:sketch_interface}
\end{figure}

We now consider an activity profile corresponding to a sharp interface separating a region of high activity $\alpha_0$ from a region of low activity $\alpha_1$. Isolated $\pm 1/2$ defects are situated at a distance $x_v$  from the interface in the region of high activity, $\alpha_0$, as illustrated in Fig.~(\ref{fig:sketch_interface}). The activity profile across this interface is given by the Heaviside step function
\[\alpha(\fet r) = \alpha_0 - \Delta \alpha H(x-x_v),\]
corresponding to a singular activity gradient $\partial_x \alpha = -\Delta \alpha \delta(x -x_v)$ with  $\Delta \alpha = \alpha_0 -\alpha_1$ the interfacial jump in activity.
An active/passive interface corresponds to $\alpha_1 =0$ and  $\Delta \alpha = \alpha_0$. In this case, we find that the self-propulsion of the $+1/2$ defect is reduced as the defect approaches the interface.
The vorticity-induced active torque tends to reorient the $\pm 1/2$ defects moving toward the interface to preferred orientations that depend on extensile/contractile activity. The $-1/2$ defect that already has the selected orientation is attracted to the wall, while that with different polarizations might be repelled.

\subsection{$+1/2$ defect}
The active force field induced by a $+1/2$ defect located at a distance $x_v$  from a sharp interface is given by  
\bea
 \mathbf F^I_+(\fet r) &=& -\Delta\alpha\delta(x-x_v) \left( \cos(2\theta_0)\fet{\hat r} - \sin(2\theta_0) \fet{\hat r}^\perp\right)\label{eq:FI_positive_interface}\;,\\
  \mathbf F^B_+(\fet r) &=& \left(\frac{ \alpha_0}{r} -\frac{\Delta\alpha}{r}H(x-x_v)\right) \mathbf{\hat e}^+\;.
  \label{eq:FB_positive_interface}
\eea
Inserting these expressions in Eq.~(\ref{eq:defect_velocity}),  we obtain the contributions to the self-propulsion velocity from interfacial and bulk active forces as 
\bea
    \mathbf v^B_+ &=& \frac{\alpha_0 \pi}{4\zeta} \mathbf{\hat e}^+ -\frac{\Delta \alpha}{2\pi\zeta^2} \mathbf{\hat e}^+\int d\mathbf r K_0\left( \frac{r}{\zeta}\right) H(x-x_v) \frac{1}{r}\;,\\
    \mathbf v_+^I &=& -\frac{\Delta \alpha}{2\pi\zeta^2} \mathbf{\hat e}^+\int_{-\infty}^\infty dy K_0\left(\frac{\sqrt{x_v^2 +y^2}}{\zeta}\right)  \frac{x_v }{\sqrt{x_v^2 +y^2}}\;.
\eea
The first term in the bulk contribution is the well-known constant self-propulsion velocity from a constant activity $\alpha_0$~\cite{ronning2022flow}. The second term is the additional drift due to the activity jump $\Delta\alpha$ and depends on the distance $x_v$ from the interface. As we will see below, this contribution suppresses the defect self-propulsion near the interface.  

\begin{figure}
    \centering
    \includegraphics[width=0.45\textwidth]{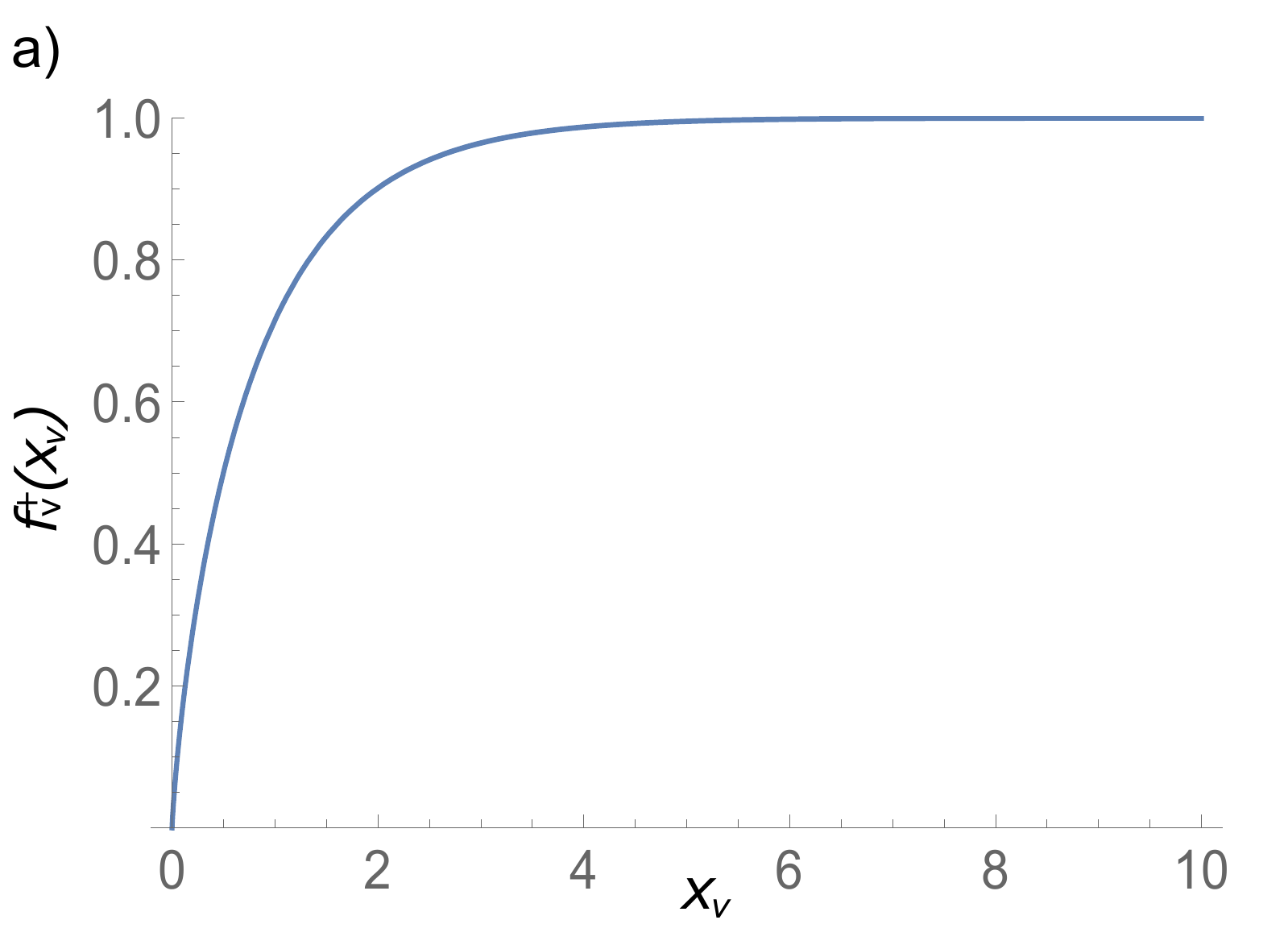}
    \includegraphics[width=0.45\textwidth]{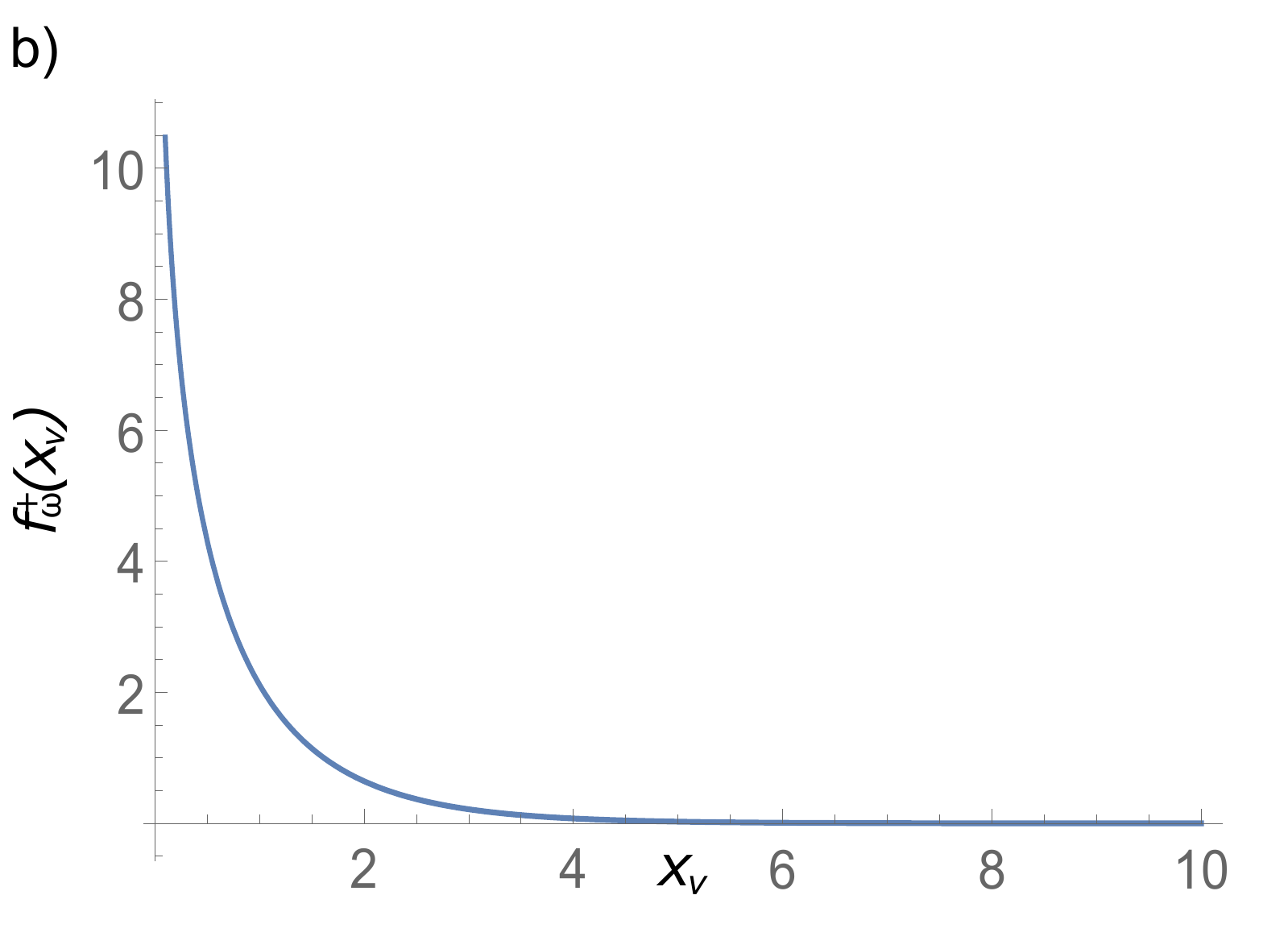}
    \caption{Plot of  (a) $f_v^+$  and (b) $f_\omega^+$ as functions of the distance $x_v$ of the $+1/2$ defect from the interface. Notice that $f_\omega^+$ diverges at $x_v=0$ due to the bulk terms. }
    \label{fig:positive_vs_wall}
\end{figure}

If we now specialize  to the case of an active/passive interface, i.e., $\Delta \alpha = \alpha_0$. In dimensional units, the self-propulsion velocity of the $+1/2$ defect is then given by 
\begin{equation}\label{eq:wall_vel_no_pressure}
     \mathbf v_+ = \frac{\alpha_0}{4\eta}\pi \ell_d f_v^+(x_v) \mathbf{\hat e}^+\;,
\end{equation}
with
\bea
f^+_v(x_v)= 1-\frac{2}{\pi^2}\int_{-\infty}^\infty d y\Bigg[ K_0\left(\sqrt{ x_v^2 + y^2}\right)  \frac{x_v }{\sqrt{ x_v^2 + y^2}} +\int_{x_v}^\infty dx K_0\left( r\right) \frac{1}{r}\Bigg]\;. 
\eea
The function $f^+_v(x_v)$  is  plotted in Fig.~(\ref{fig:positive_vs_wall} a). 
We note that the self-propulsion speed vanishes as the defect hits the interface $x_v =0$. In other words, the defect slows down as it approaches the interface, and eventually remains at rest at the interface. We note that  Eq. ~(\ref{eq:wall_vel_no_pressure}) is obtained by incorporating the incompressibility constraint only in the $\mathbf v_+^0$ term. Additional pressure gradients may arise due to activity jump. These are however, difficult to obtain analytically and are not included in this study.

We now compute the vorticity at the defect position to investigate how its contribution to the active torque tends to re-orient the defect as it approaches the interface to a stable orientation. From Eq.~(\ref{eq:defect_vorticity}), we obtain the following expressions for interfacial and bulk contributions 
\bea
    &\omega^I_+ = -\frac{\Delta \alpha}{2\pi \zeta^2}\sin(2\theta_0) \int_{-\infty}^\infty dy\left[ K_0\left(\frac{\sqrt{x_v^2 +y^2}}{\zeta}\right)  \frac{x_v^2}{(x_v^2 +y^2)^{3/2}} +  K_1\left(\frac{\sqrt{x_v^2 +y^2}}{\zeta}\right)\frac{x_v^2}{\zeta (x_v^2 +y^2)} \right]\;,\\
    & \omega^B_+ = -\frac{\Delta \alpha}{2\pi \zeta^2} \sin{2\theta_0} \int_{-\infty}^\infty dy\left[ K_0\left(\frac{\sqrt{x_v^2 +y^2}}{\zeta}\right) \frac{1}{\sqrt{x_v^2 +y^2}}  -\int_{x_v}^\infty dx K_0\left(\frac{r}{\zeta}\right) \frac{x}{r^3}\right]\;,
\eea
where the bulk vorticity diverges at $x_v = 0$. The total defect angular velocity is given by the sum of these two contributions evaluated at the defect core. In dimensional units, it is given by
\beq
 \omega_+(x_v,\theta_0) = -\frac{\Delta \alpha}{2\pi \eta}\sin(2\theta_0) f_\omega^+(x_v)\;,
\eeq
where the wall-dependence function $f_\omega^+(x_v)$ 
is plotted in Fig.~(\ref{fig:positive_vs_wall} b).
Hence, near an active/passive interface, the vorticity-induced rotation is at a rate $\dot\theta_0=\frac{1}{2}\omega_+(x_v,\theta_0)$ until $\omega_+(x_v,\theta_0)=0$. It is clear from Fig.~(\ref{fig:positive_vs_wall} b) that re-orientation only occurs within a distance of order $\ell_d$ from the wall. As the $+1/2$ defect approaches the wall, $f_\omega^+$ increases  and eventually diverges at $x_v\rightarrow 0$. This means that the defect tends to re-orient its polarization until $\sin(2\theta_0)=0$. From the stability criterion that $\frac{d\omega_+}{d\theta_0}<0$, this corresponds to the stable orientation $2\theta_0 = \pi$ for $\alpha_0<0$ (extensile) and $2\theta_0 = 0$ for $\alpha_0>0$ (contractile). In both cases, the defect polarization is normal to the interface $\mathbf e_+ = [\mp 1,0]$ and points away from the interface for extensile systems and into the interface for contractile systems, respectively. 
\jr{}
Numerical simulations \cite{zhang2021spatiotemporal} report that $+1/2$ defects tend to reorient and drift parallel to the boundary when the angle between the interface and the incoming velocity is below a critical value that depends on activity. 
Above this critical angle, i.e more head-on collisions, the defect hits the wall and tunnels through it.
This effect is likely coming from the additional contributions to the active torque that are not considered here, namely the interactions between defects, deformations in the nematic order parameter due to the wall and the coupling to flow alignment.
It is likely that these terms are important close to the interface, both for determining the defect orientation and the tunneling effect observed both experimentally and numerically ~\cite{zhang2021spatiotemporal}.

\subsection{$-1/2$ defect}
\begin{figure}[t]
    \centering
    \includegraphics[width = 0.45 \textwidth]{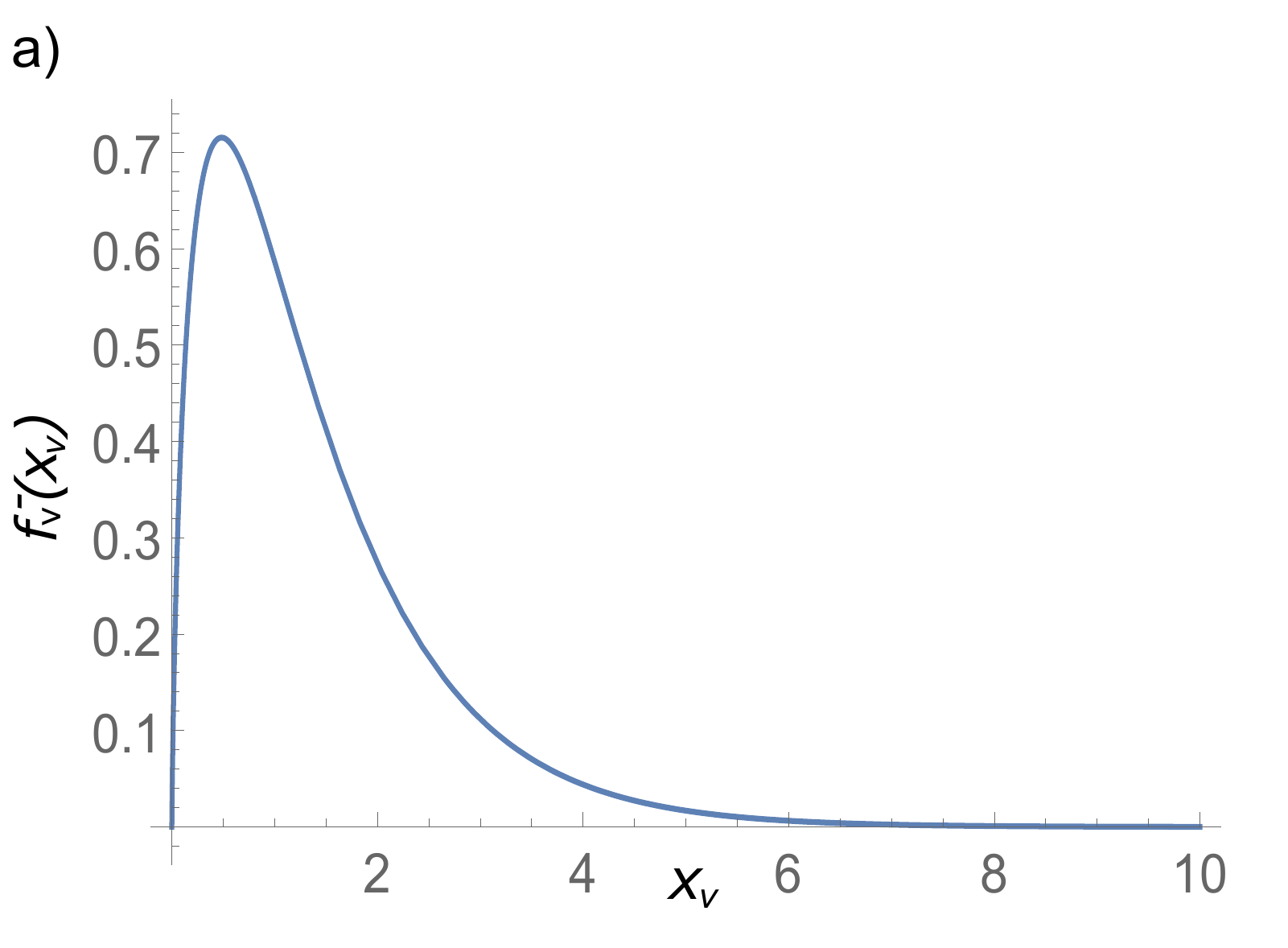}
    \includegraphics[width = 0.45 \textwidth]{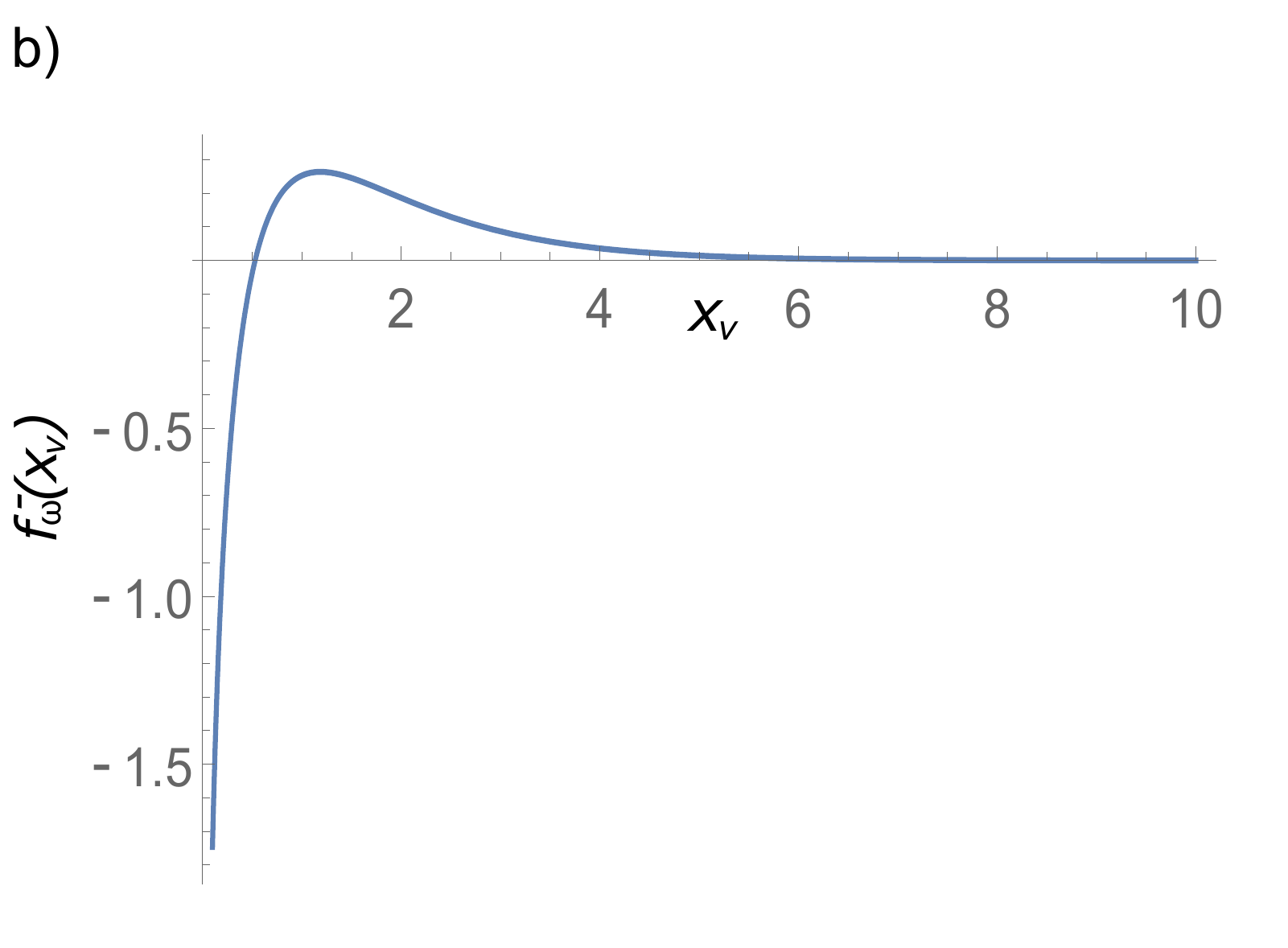}
  \caption{Profile of (a) $f_v^-(x_v)$ and (b) $f_\omega^-(x_v)$ as function of $x_v$. Note that the function $f_\omega^-(x_v)$ diverges at $x_v=0$. }
    \label{fig:wall_dependence_negative}
\end{figure}

The components of the interfacial active force due to a $-1/2$ defect at a distance $x_v$ from the activity jump are given by 
\bea
 F^I_{x-} = -\frac{\Delta \alpha}{r} \delta(x-x_v) (x\cos{2\theta_0} +y\sin{2\theta_0})\;, \\
 F^I_{y-} = -\frac{\Delta \alpha}{r} \delta(x-x_v) (-y\cos{2\theta_0} +x\sin{2\theta_0})\;. 
\eea
The corresponding bulk active force is 
\bea
   F^B_{x-} = (\alpha_0 - \Delta\alpha H(x-x_v))\frac{1}{r^3}
    [(y^2 -x^2)\cos{2\theta_0} -2xy\sin{2\theta_0}] \;,\\
   F^B_{y-} = (\alpha_0 - \Delta\alpha H(x-x_v))\frac{1}{r^3}
   [(y^2 -x^2)\sin{2\theta_0} +2xy\cos{2\theta_0}]\;.
\eea
Using these expressions, and  neglecting the  contribution from the pressure gradient, the net drift velocity of the defect can be written as 
\bea \label{eq:negative_wall_force}
\mathbf v_-(x_v) = -\frac{\Delta \alpha}{2\pi\eta}\ell_d  f_v^-(x_v) \mathbf{\hat n}^- ,
\eea
where $\mathbf{\hat n}^- = \cos(2\theta_0) \mathbf{\hat x} +  \sin(2\theta_0) \mathbf{\hat y}$. The function $f_v^-(x_v)$ describes the dependence on the distance $x_v$ to the interface  
and is given by 
\[f_v^-(x_v) = \int_{\infty}^\infty d y K_0(\sqrt{x_v^2 +y^2}) \frac{x_v}{\sqrt{x_v^2 +y^2}}- \int_{x_v}^\infty\int_{-\infty}^\infty dxdy K_0(r) \frac{x^2 -y^2}{r^3}\;.\]
It has been be evaluated numerically and is plotted in Fig. (\ref{fig:wall_dependence_negative} a). The $-1/2$ defect acquires a finite self-propulsion close to the wall in a region of thickness of order $\ell_d$ near the activity jump. Its motion is either towards or away from the boundary, depending on the defect's orientation
and the sign of the activity.

To see how the $-1/2$ reorients as it approaches the interface, we evaluate the flow vorticity at the defect core as a function to the wall distance. Again, there are contributions to the vorticity from both flows driven by interfacial and bulk forces, given by
\begin{align}
\omega^I_- =& -\frac{\Delta \alpha}{2\pi \zeta^2}\sin{2\theta_0} \int_{-\infty}^\infty dy \left[K_1\left(\frac{r}{\zeta}\right) 
\frac{x_v^2}{\zeta(x_v^2 +y^2)}- K_0\left(\frac{\sqrt{x_v^2 +y^2}}{\zeta}\right) \frac{x_v^2}{\sqrt{x_v^2 +y^2}^3} \right]\;,\\
\omega^B_- =& -\frac{\Delta \alpha}{2\pi \zeta^2} \sin{2\theta_0} \int d\mathbf r K_0\left(\frac{r}{\zeta}\right)
\Bigg(\delta(x-x_v) \frac{y^2 -x^2}{r^3} 
+H(x-x_v) \frac{3x(x^2 -3y^2)}{r^5}
 \Bigg)\;.
\end{align}
Note that the bulk term diverge when $x_v \rightarrow 0$.
The total angular velocity of the $-1/2$ defect can then be written as
\bea
 \omega_- (x_v,\theta_0)= -\frac{\alpha_0}{2\pi \eta}\sin(2\theta_0) f^-_\omega(x_v)\;. 
\eea
The function $f_\omega(x_v)$ has been calculated numerically and is shown in Fig.~(\ref{fig:wall_dependence_negative} b)). The dependence on the wall distance $x_v$ changes sign near the wall, indicating that the vorticity tends to rotate the defect to a preferred orientation at the wall. The preferred orientation is determined by the stationary condition $\sin{2\theta_0}=0$, and the stability criterion $\frac{d\omega_-}{d\theta_0}<0$, which implies that $\theta_0 = 0$ for $\alpha_0<0$ and $2\theta_0 = \pi$ for $\alpha_0>0$. In other words for extensile activity the stable orientation of a $-1/2$ defect at a sharp active/passive interface corresponds to a polarization $\mathbf e_{-} = [1, 0]$.
Therefore, as a result of both their self-induced translational and rotational motion, in an extensile system $-1/2$ defects are attracted to a sharp active/passive interface and orient themselves with one of the three  axis normal to the interface. This is consistent with the accumulation of negative topological charge observed in experiments at active/passive interfaces~\cite{thijssen2021submersed,zhang2021spatiotemporal} and near physical walls~\cite{hardouin2020active}, as well as in simulations~\cite{LSthesis,scharrer2022spatial}.

\section{Conclusion}
\label{sec:conclusion}
Activity gradients or sharp jumps can guide the motion and orientation of nematic defects. In a constant activity gradient, $+1/2$ defects acquire an angular velocity that may rotate their orientation  such that the defect polarization aligns parallel to the activity gradient. The defects then self-propels in the direction of the gradient, always moving towards regions of lower magnitude of activity, where it is less motile. 
Thus, we expect that activity gradients will introduce more circular motion in the trajectories of the $+1/2$ defects. 
In contrast, a constant activity gradient yields  no net vorticity or active force at the core of the $-1/2$ defect, which remains stationary. 

 We find that the self-propulsion velocity of $+1/2$ defects moving towards a sharp active/passive interface is also reduced, and that the defect will eventually stagnate at the wall.
 In contrast, $-1/2$ defects acquire a finite propulsion speed in the interfacial region and can overcome the positive defects, explaining the observation of negative charge accumulation in experiments and simulations~\cite{thijssen2021submersed,zhang2021spatiotemporal,hardouin2020active,LSthesis,scharrer2022spatial}.
 We also predict that the active torque acting on a $+1/2$ defect that reaches the interface tends to reorient it toward a preferred polarization that is perpendicular to the interface and points away/toward it depending on extensile/contractile activity. The vorticity-induced active torque also acts on the orientation of a $-1/2$ defect migrating toward interface, by rotating the defect until it reaches the stable orientation which minimizes the net vorticity at the defect position. We show that a $-1/2$ defect with a stable orientation gets attracted to a sharp interface. This stable orientation is selected by the sign of activity, i.e whether the system is contractile or extensile. 
Tunneling across the interface observed numerically may be due to soft interfaces where the activity gradients are not sufficiently steep, as well as due to defect interactions and other hydrodynamic effects. Here, we have neglected additional contributions of pressure gradients induced by activity gradients, as well as elastic stresses, flow alignment, nematic distortions due to the active/passive interface and defect interactions, which may change qualitatively the defect dynamics. 

Our results offer a simple understanding of the dynamics of nematic defects in the presence of spatially-varying activity. They can provide the starting point for designing structures capable of controlling defect dynamics and associated active flows.\\


\enlargethispage{20pt}

\ethics{No ethical dilemmas where encountered in the preparation of this paper. }

\dataccess{This is primarily theoretical work and does not have any experimental data. The computational data and codes for  FEniCS are available on GitHub: https://github.com/jonasron/Defect-Flows}

\aucontribute{J.R. derived the analytical results and performed finite-element
simulations for finite domains. L.A. verified all analytical calculations. All authors contributed
to a critical discussion of the analytical and numerical results and participated in writing the manuscript.}

\competing{We declare we have no competing interests.}

\funding{M.C.M. was supported by the US National Science Foundation Grant No. DMR-2041459. J.R. and L.A. acknowledge support from the Research Council of Norway through the Center of Excellence
funding scheme, Project No. 262644 (PoreLab).}

\ack{M.C.M. thanks Mark Bowick, Luca Scharrer and Suraj Shankar for illuminating discussions. L.A. and J.R. are thankful to Jorge Vi\~nals for simulating discussions.}


\bibliographystyle{RS}
\bibliography{refs}

\end{document}